\documentclass[11pt,superscriptaddress, reprint, pre, longbibliography]{revtex4-1}

\usepackage[utf8]{inputenc}
\usepackage{amsmath}
\usepackage[utf8]{inputenc} 
\usepackage[english]{babel}
\usepackage{tikz}
\usepackage{amsmath}
\usepackage{graphicx}
\usepackage[colorlinks=true, allcolors=blue]{hyperref}
\usepackage{amsfonts}

\newcommand{\Tau}{\tau}

\begin{document}

\author{Nathan Abitbol}
\affiliation{
Universit\'e  Paris-Saclay, CNRS,  FAST,  91405,  Orsay, France
}

\author{Alex Hansen}
\affiliation{
PoreLab, Department of Physics,\\ Norwegian University of
Science and Technology, N-7491, Trondheim, Norway
}
\author{Alberto Rosso}
\affiliation{
Universit\'e  Paris-Saclay,  CNRS,  LPTMS,  91405,  Orsay, France
}

\author{Laurent Talon}
\email{laurent.talon@universite-paris-saclay.fr}
\affiliation{
Universit\'e  Paris-Saclay, CNRS,  FAST,  91405,  Orsay, France
}

\begin{abstract}
We study the flow of a Bingham yield-stress fluid in a pore network model where the throats have radii drawn from a uniform distribution. We consider the case in which a fraction of the largest radii is blocked. The fluid can flow only through the percolating cluster that exists when the fraction is above the percolation threshold. Two distinct flow regimes are identified: above the percolation threshold the flow curve can be characterized by deterministic values of the critical pressure drop, permeability, and other observables, with subleading fluctuations that we quantify. At the percolation threshold these quantities become non-self-averaging, and their scaling is governed exclusively by the critical percolation backbone, independent of the specific realization of the radii.
\end{abstract}

\title{Flow of yield stress fluid in a percolating network}

\maketitle

\section{Introduction}

Yield-stress fluid flows in porous media have numerous everyday and industrial applications~\cite{bird87,coussot05}. 
For instance, grouts are injected into soils for stabilization~\cite{seiphoori22}, while foams can be injected to decontaminate or to limit the propagation of pollutants~\cite{portois18}. Some heavy oils in the ground might also exhibit a yield stress depending on temperature~\cite{ghannam12}. 
Owing to their ubiquity, such problems have been extensively studied~\cite{pascal81,al-fariss87,lopez03,balhoff04,talon13b,waisbord19,fraggedakis21,pourzahedi24}, with a predominant focus on \emph{saturated} porous media. 
In this case, Darcy’s law is modified for yield stress fluids. First, the flow initiates only when the applied pressure drop exceeds a critical threshold, $\Delta P_0$.
As shown in references \cite{roux87,talon13b}, the critical threshold results from a path minimisation principle, and above this threshold, the flow rate scales non-linearly with $\Delta P$ due to the pressure-dependent activation of flow pathways.
This behavior manifests as an effective permeability that varies with $\Delta P$ but asymptotically approaches the classical Newtonian permeability at high pressures, provided the system exceeds the Representative Elementary Volume (REV) scale.
On the other hand, $\Delta P_0$ exhibits a linear dependence with the system size $L$~\cite{liu19,fraggedakis21,chaparian24}, implying that $\Delta P_0 / L$ also converges to a constant for sufficiently large systems.
In our previous studies, we mainly used a ``directed'' pore network model~\cite{liu19}, which assumes the direction of the flow. This allows for faster computation and a better characterization of the flow, for example, on a Cayley tree~\cite{schimmenti23,munier2025}.
Studies of unsaturated porous media with yield-stress fluids are more scarce. Chen \emph{et al.}~\cite{chen05} introduced a pore network model with either dynamic or static bubbles, while Pourzahedi and Frigaard~\cite{pourzahedi24} used a dynamic bubble approach. The latter assumed the direction of the flow in the links.

In this study, we aim to generalise previous analyses to unsaturated porous media, where the pore space comprises a yield-stress fluid and a non-wetting phase (e.g. air bubbles). To model this system, we assume that the non-wetting phase occupies the largest pore throats. We also consider these non-wetting phases to be trapped, which blocks the pores and makes them unable to conduct flow.
Consequently, these blocked links can be considered absent from the original network, corresponding to a standard percolation process.
As both the critical pressure threshold and the flow are the result of a minimisation process, we expect that limiting the number and geometry of potential flow paths will have a significant impact on the flow properties.

Following the frameworks of Wilkinson~\cite{wilkinson84} and Lenormand and Zarcone~\cite{lenormand85}, but more relevantly of Chen \emph{et al.}~\cite{chen05}, on percolation-based invasion models, we assume that a pore throat becomes blocked if its diameter exceeds the critical value set by the capillary pressure. The non-wetting fluid saturation is then related to the fraction $1-p$ of removed links. It is important to note, however, that we use a non-directed pore network model, \emph{i.e} without assuming the direction of the flow in each link. As it will be discussed later, this is a crucial assumption for high non-wetting saturation where the flow becomes very tortuous and cannot be imposed \emph{a priori}.

We investigate the flow properties of a yield-stress fluid as a function of $p$. 
In particular, we show that, near the percolation threshold $p_c$, both the permeability and the apparent pressure threshold scale with the system size following non-trivial exponents. 
Another important finding near the percolation point is the loss of self-averaging for these quantities: the standard deviation of the measured values no longer decays faster than their mean as the system size increases. 
Consequently, these quantities do not converge to a deterministic value in the large-system limit (e.g., in the representative elementary volume sense), but instead remain inherently stochastic.

The paper is organized as follows. 
We first describe the model and the governing equations. 
Section~\ref{sec:low_Q} examines the flow properties in the low-flow-rate regime, while Section~\ref{sec:high_Q} focuses on the high-flow-rate behavior. 
Finally, in Section~\ref{sec:hom}, we discuss the relative roles of pore-size disorder and the geometry of the percolation cluster.

\section{\label{sec:model}Model}

In this work, we employ a pore network model \cite{sochi08, pascal81}, a simplified representation of a porous medium in which nodes correspond to pores and links represent pore throats. We consider a two-dimensional diamond network of width $W$ and length $L$. A pressure $P_i$ is defined at each node $i$. Each link $(i,j)$ is modeled as a cylindrical conduit of length $l$ and a radius $r_{ij}$ uniformly distributed between between $r_{\mathrm{min}}$ and $r_{\mathrm{max}}$. A global pressure drop $\Delta P$ is imposed at the boundaries along $L$: $P_{\mathrm{in}} = \Delta P$ at the inlet nodes and $P_{\mathrm{out}} = 0$ at the outlet nodes.

\begin{figure}[ht]
    \centering
    \includegraphics[width=0.9\hsize]{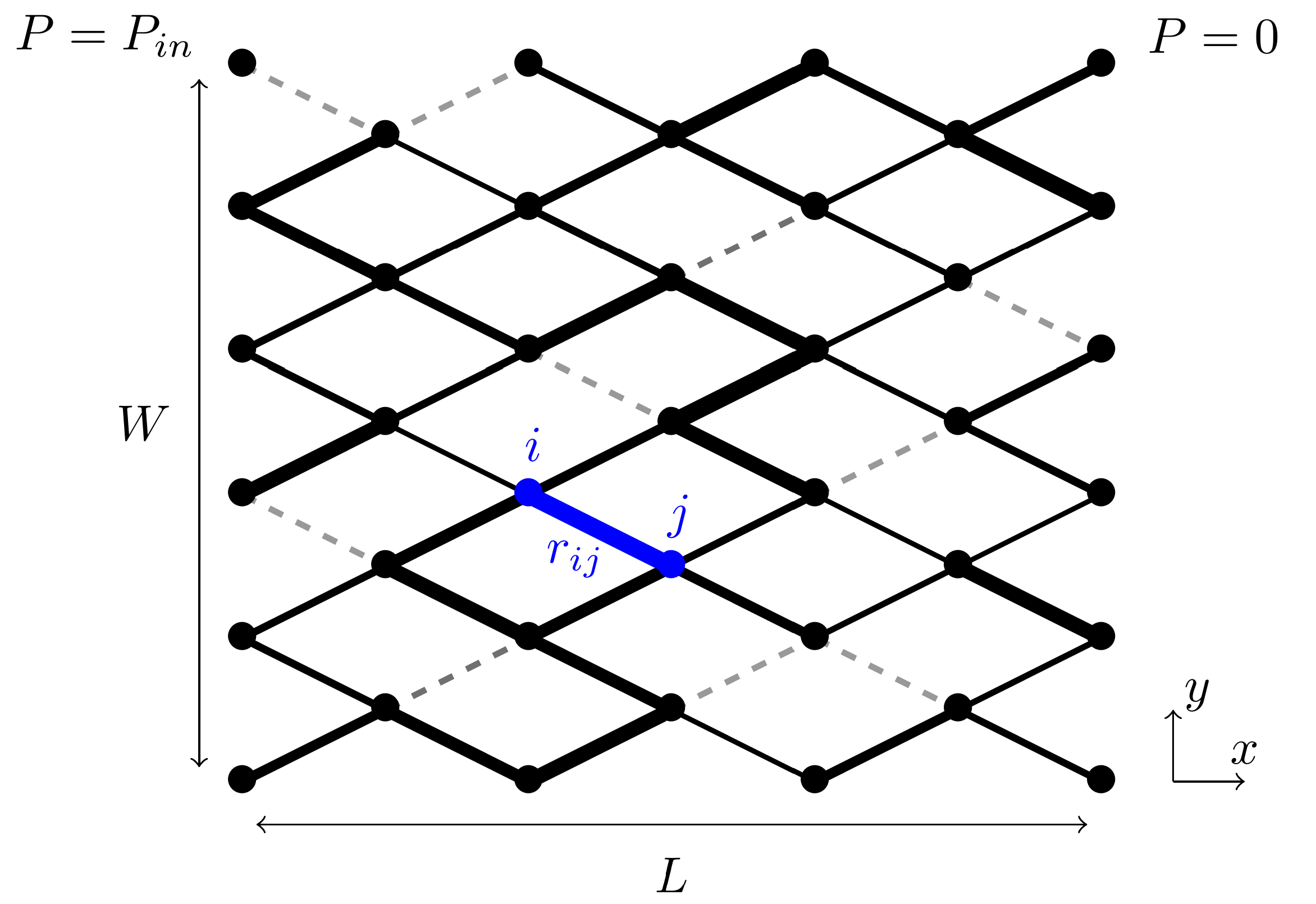}
    \caption{Sketch of a pore network of size $L$ and width $W$. The bold links indicate a larger radius, and cut links are represented by dashed lines. The pressure $P = P_{in}>0$ is imposed at the inlet (left boundary), while $P = 0$ is imposed at the outlet (right boundary). Here, the flow goes from left to right.}
    \label{fig:sketch}
\end{figure}

All the flow $q_{ij}$ occurs in the pore throats (see Fig.~\ref{fig:sketch}) and the conservation of mass at each node $i$ yields the Kirchhoff law:
\begin{equation}
    \sum_{j \in V(i)} q_{ij} = 0,
    \label{khirchoff}
\end{equation}
where $V(i)$ denotes the set of neighbors of node $i$. For each conducting link, we approximate the flow rate of a Bingham fluid, with $\Tau_c$ the yield stress, using a piecewise-linear profile:
\begin{equation}
    q_{ij} =
     \begin{cases}
       \sigma_{ij} \left( \delta P_{ij} - \Tau_{ij} \right) & \text{if } \delta P_{ij} > \Tau_{ij}, \\
       0 & \text{if } |\delta P_{ij}| \leq \Tau_{ij}, \\
       \sigma_{ij} \left( \delta P_{ij} + \Tau_{ij} \right) & \text{if } \delta P_{ij} < -\Tau_{ij},
     \end{cases}
     \label{bingham_link_flow}
\end{equation}
where $\sigma_{ij} = \pi r_{ij}^4 / (8\eta)$ is the hydraulic conductivity, $\delta P_{ij} = P_i - P_j$ the pressure difference, and
\begin{equation}
    \Tau_{ij} = \frac{2 \Tau_c l_{ij}}{r_{ij}}
\end{equation}
the local yield pressure  threshold.

This piecewise-linear profile neglects the nonlinear region near the threshold $\Tau_{ij}$ that arises from the full Poiseuille-like solution for a Bingham fluid \cite{buckingham21, chen05}.
Nonetheless, we expect this approximation to preserve the correct macroscopic scaling behavior. Despite its simplicity, this minimal model is particularly well suited for evaluating flow statistics and associated quantities such as the effective permeability. It retains the essential scaling laws observed in more sophisticated numerical methods, including lattice Boltzmann \cite{talon13b} and finite element simulations \cite{fraggedakis21}.

In this work, we study the case where a fraction $1-p$ of the throats corresponding to the largest radii is blocked due to the presence of droplets. In particular, it is interesting to characterize the flow  at the percolation threshold $p = p_c$.
The Newman-Ziff algorithm \cite{newman01} is used to compute the percolation threshold $p_c(L)$ for each realization. Starting from a fully blocked network with pre-assigned radii $r_{ij}$, links are progressively opened in increasing order of radius. That is, the links with the  smallest radii are added first, and smaller ones are subsequently included until a spanning cluster connecting the inlet to the outlet emerges. The determination of spanning clusters does not depend on the flow solution, but only on whether a continuous path of un-obstructed links exists from inlet to the outlet. Clusters that are geometrically disconnected from either boundary do not contribute to the flow.

\subsection*{Generalized Darcy equation}

For a given given configuration of blocked throats and assigned radius we can study the flow rate as a function of the  pressure drop $\Delta P$. It obeys a generalized version of the Darcy law, namely:
\begin{equation}
    Q = \kappa_\mathrm{eff}(\Delta P) W \left[ \frac{\Delta P}{L} - \frac{\Delta P^*(\Delta P)}{L} \right].
    \label{eq:nonlinear_darcy}
\end{equation}
Here $W$ is the system's width, $\kappa_\mathrm{eff}(\Delta P)$ is the effective permeability and $\Delta P^*(\Delta P)$ the apparent critical pressure. Both quantities  are piece-wise constant as a function of $\Delta P$: they remain unchanged as long as the number of open links does not vary, and jump to higher values when new links open. This dependence on $\Delta P$ is the origin of the non-linearity in Darcy's law. 
In Appendix \ref{app:darcy}, we show that these quantities can be written as:
\begin{align}
    \kappa_\mathrm{eff}(\Delta P) &= \frac{L Q^2 }{W} \left( \sum_{\langle ij \rangle} \frac{1}{\sigma_{ij}} q_{ij}^2 \right)^{-1}, \label{eq:kappa_def}\\
    \Delta P^*(\Delta P) &= \frac{1}{Q} \sum_{\langle ij \rangle} \Tau_{ij} |q_{ij}|.
    \label{eq:Pstar_def}
\end{align}
where the sum is taken over the open links.

\subsection*{Flow regimes and effective permeability}

For yield stress fluids, the behavior of the effective permeability $\kappa_{\mathrm{eff}}$ as a function of the imposed pressure drop $\Delta P$ can be naturally divided into three distinct regimes:

\textbf{(i) Low flow regime} — When $\Delta P < \Delta P_0$, the effective permeability vanishes: $\kappa_{\mathrm{eff}} = 0$. At the threshold $\Delta P = \Delta P_0$, the first flow-carrying channel connecting the inlet to the outlet opens. This critical channel is not necessarily straight—especially at the percolation threshold $p = p_c$—and can be highly tortuous. Mathematically, the threshold pressure is obtained by solving a minimization problem over all possible paths $\mathcal{C}$:
\begin{equation}
    \label{eq:P0_minimization}
    \Delta P_0 = \min_{\mathcal{C}}  \sum_{\langle ij\rangle \in \mathcal{C}} \tau_{ij}
\end{equation}
This corresponds to a competition between selecting a short path and one composed of links with the lowest local thresholds. In practice, Dijkstra's algorithm (in its undirected form) \cite{dijkstra59} is used to identify the critical path $\mathcal{C}_0$. In this regime, where only a single channel carries flow, the link fluxes are all equal to the total flow rate $Q$, and the effective permeability at threshold is given by:
\begin{equation}
   \kappa_0 \equiv \kappa_{\mathrm{eff}}(\Delta P_0) = \left( \sum_{\langle ij \rangle \in \mathcal{C}_0} \sigma_{ij}^{-1} \right)^{-1}.
    \label{eq_permeability}
\end{equation}

\textbf{(ii) High flow regime} — In the opposite limit of large pressure drops $\Delta P \gg \Delta P_0$, all links in the network are open and conduct flow. The system behaves as a Newtonian network, and the effective permeability saturates to a finite asymptotic value:
\begin{align}
    \kappa_\infty & \equiv \lim_{\Delta P \to \infty} \kappa_{\mathrm{eff}}(\Delta P), \\
    \Delta P_\infty & \equiv \lim_{\Delta P \to \infty} \Delta P^*(\Delta P).
\end{align}

\textbf{(iii) Intermediate regime} — Between these two limits lies a nontrivial intermediate regime, where the pressure drop exceeds the threshold ($\Delta P > \Delta P_0$) but is not yet large enough to activate all links. In this regime, the number of active channels progressively increases with $\Delta P$, and the topology of the flow paths evolves. To study this behavior and solve Eqs.~(\ref{khirchoff}) and~(\ref{bingham_link_flow}), we use the Augmented Lagrangian method introduced in~\cite{talon20}, a variational approach that involves two coupled fields: one enforcing mass conservation (a linear but nonlocal constraint), and one imposing the local nonlinear rheology.

In our simulations, $l=1$, $\eta=10^{-3}$, $\tau_c=10$, and the radii $r_{ij}$ are uniformly distributed between $r_{\mathrm{min}} = 0.17$ and $r_{\mathrm{max}} = 0.32$, with $W=L$ and $L$ ranging from 32 to 1024.

The structure of the paper is as follows. In Section~3, we analyze the low flow rate regime, with particular attention to the computation of the threshold pressure $\Delta P_0$ as a function of the link density $p \in [p_c, 1]$. We then focus on the specific case of the percolation threshold $p = p_c$. Section~4 is devoted to the high flow regime, where all links are open. In Section~5, we explore the intermediate regime and introduce an effective homogenized description that captures the flow behavior at $p = p_c$.
\begin{figure*}[ht]
    \centering
    \includegraphics[width=\hsize]{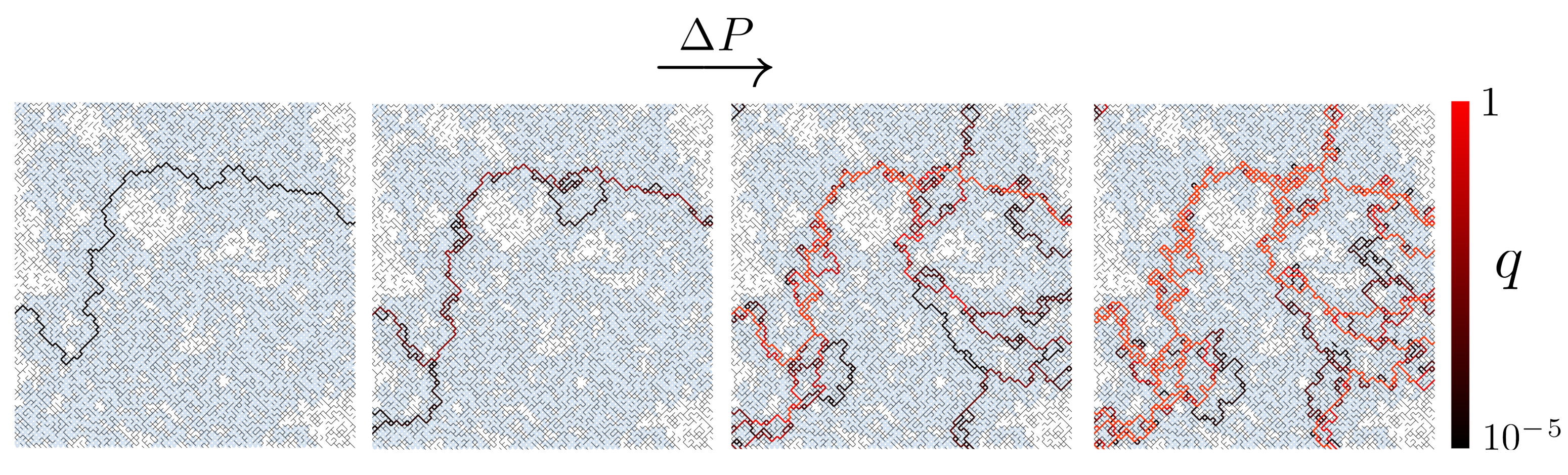}
    \caption{Snapshots of simulations on a $128 \times 128$ network at the percolation threshold $p=p_c$. The cut links in the network are represented in black. The blue nodes belong to the percolating cluster, while the white ones denote the non-spanning clusters : fluid can not flow through these. With increasing pressure drops $\Delta P$, more paths start flowing, increasing the medium's effective permeability.}
    \label{fig:image_flow}
\end{figure*}
\section{\label{sec:low_Q}Small flow rate limit $Q \rightarrow 0$}

To determine the pressure $\Delta P_0$ and the first flow-carrying channel connecting the inlet to the outlet, one must solve the minimization problem defined in Eq.~\eqref{eq:P0_minimization}. This corresponds to finding a non-directed polymer (NDP) that minimizes the total sum of pressure thresholds. In the language of polymers in random media, this sum is analogous to the lowest energy of the polymer. We begin by reviewing the main results known in the context of polymers in random media.

\subsection{Polymers in random media: state of the art}

 Most existing results have been obtained within the directed polymer (DP) approximation, where all paths are constrained to have the same minimal length $L$, and $p=1$. The optimization over pressure thresholds leads to the following expression~\cite{kpz86}:
\begin{equation}
\Delta P_0^{DP} = c^{DP} L + \chi^{DP} L^{1/3}
\label{eq:KPZ}
\end{equation}
In this expression, the first term is deterministic and extensive, while the second term is stochastic. The random variable $ \chi^{DP} $, of order unity, is distributed according to the Tracy–Widom family of distributions. Kardar and Zhang \cite{kardar87} showed that such directed polymers belong to the KPZ universality class, as the height $h(x,t)$ of a growing interface can be mapped to the ground state energy of a polymer of length $t$ ending at position $x$, at zero temperature.
Note that $c^{DP}$ represents the average pressure threshold along the critical channel. Due to the optimization process, we have $ c^{DP} \ll \langle \tau \rangle $, where $\langle . \rangle$ denotes the average over the radii distribution.

The non-directed case for $ p = 1 $ has been previously studied in Refs.~\cite{schwartz98,marsili98} using the same non-directed Dijkstra algorithm employed in this work. These studies demonstrated that the problem belongs to the same universality class as the directed polymer, and in particular, that Eq.~\eqref{eq:KPZ} still holds in this case. Variants involving correlated disorder have been explored in \cite{schorr03}, as well as versions with power-law distributed weights \cite{hansen04,halpinhealy98}.
For $p<1$, the problem was first analyzed within the directed polymer approximation in Ref.~\cite{balents92}. They found that the fluctuations of the ground state energy grow as $ L^{1/3} $, as in the case $ p = 1 $, for all $ p > p_d \simeq 0.6445 $, where $ p_d $ is the directed percolation threshold. At this critical point, the fluctuations increase as $ L^{1/2} $. This can be understood from the fact that the number of available directed paths becomes very limited, suppressing the possibility of optimization. As a consequence, the pressure thresholds along the critical path are distributed similarly to those in the bulk of the network, and the $ L^{1/2} $ scaling of the fluctuations results from the central limit theorem, reflecting the sum of $ L $ independent random variables.

In Ref.~\cite{seno97}, Seno \emph{et al.} investigated the ground state of the non-directed polymer in a disordered medium for \( p \in [p_c,1] \). Their analysis relied on the transfer matrix method introduced by Derrida~\cite{derrida88}, which enables exact enumeration of paths on samples of size \( L \times W \), with a modest transverse width \( W < 9 \), and binary disorder. Based on extrapolations from these finite-width systems, they concluded that for \( p > p_c \), the length of the ground state path remains (in our notations) proportional to \( L \), and the corresponding ground state energy is extensive. 

At the percolation threshold \( p = p_c \), however, they identified the ground state path with the chemical path, i.e., the shortest path on a critical percolation cluster. This path exhibits a fractal geometry characterized by a scaling \( \ell_{\mathrm{chem}} \propto L^{D_\mathrm{min}} \), with a fractal dimension \( D_\mathrm{min} \simeq 1.13 \)~\cite{barma85,havlin84,lee99}. Consequently, the ground state energy scales as \( L^{D_\mathrm{min}} \) at criticality.

Due to the limitations of the transfer matrix approach, particularly the restricted system width, their study could not provide conclusive results regarding the fluctuations of the ground state energy—especially in the intermediate regime \( p \in [p_c, p_d] \) and at the critical point \( p = p_c \).

\begin{figure*}[ht]
    \centering
    \includegraphics[width=0.4\hsize]{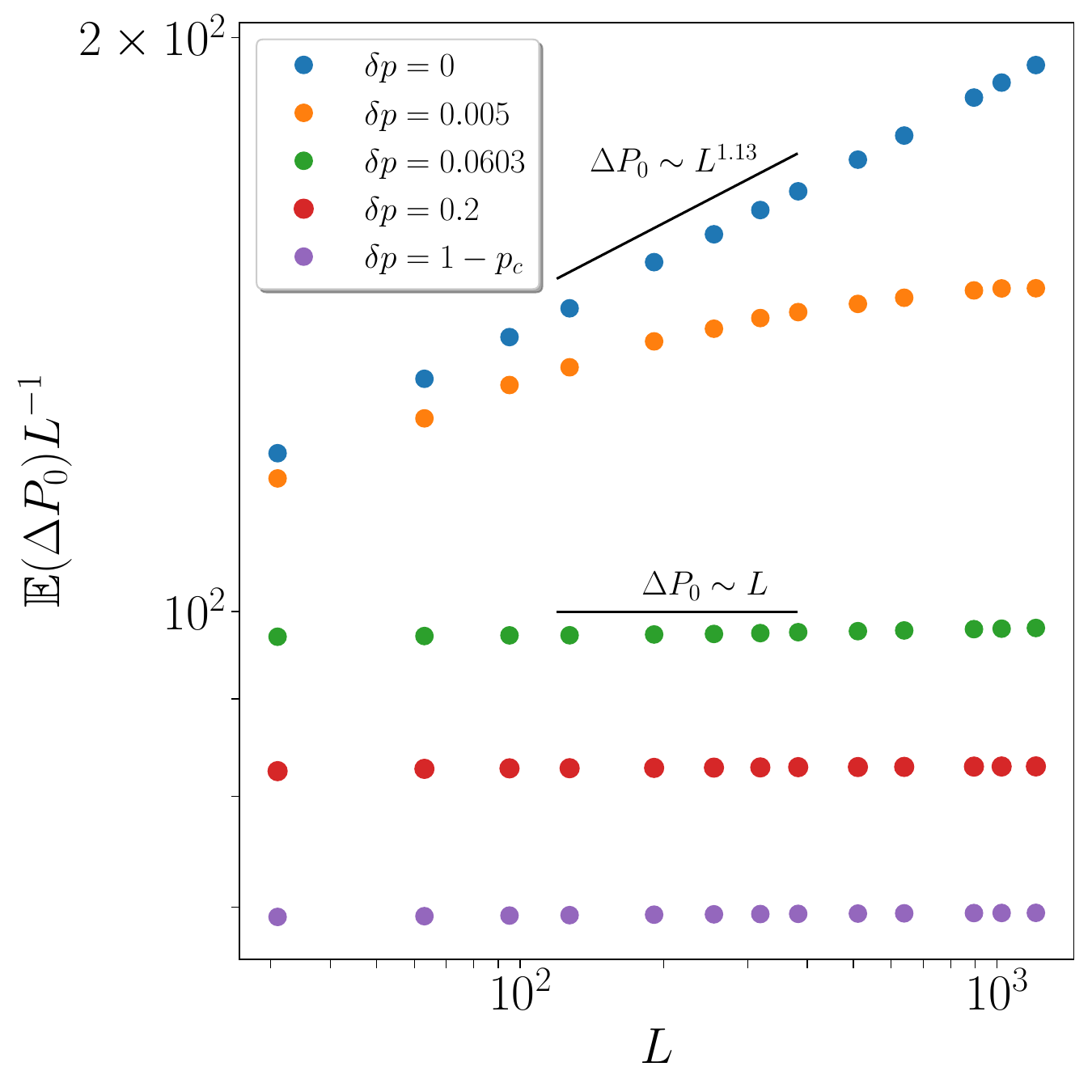}
    \includegraphics[width=0.4\hsize]{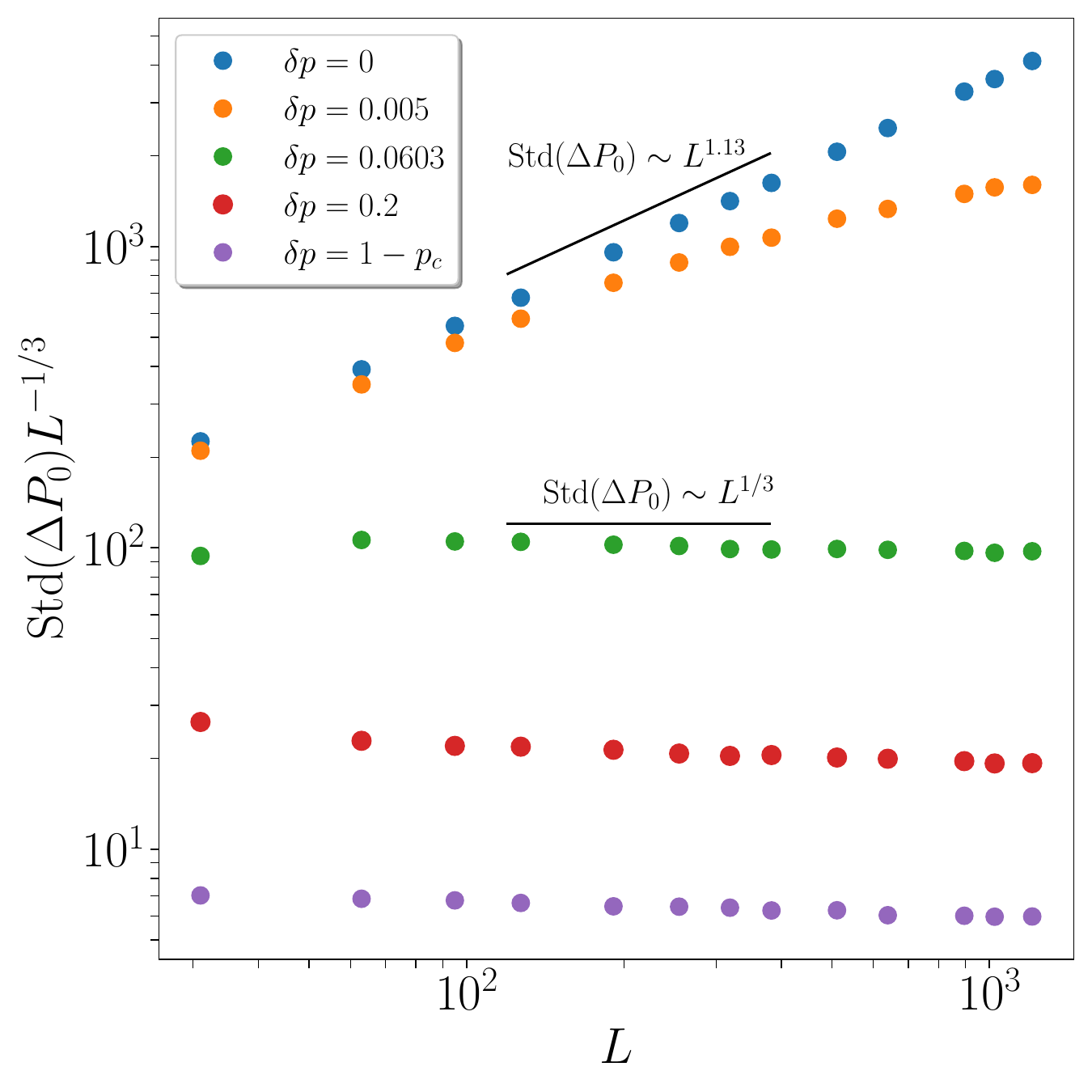}
    \caption{
    \textbf{Left:} Average pressure threshold \( \Delta P_0 \) as a function of the system size \( L \), at fixed distance from the percolation threshold \( \delta p = p - p_c \). 
        \textbf{Right:} Standard deviation of \( \Delta P_0 \) versus \( L \).
        Averages are performed over 10 000 disorder realizations.
    }
    \label{fig:Energy_Q0}
\end{figure*}

\subsection{Non-directed Polymers in random media: $\Delta P_0$}
In the present work, we overcome these limitations by employing Dijkstra's algorithm, which allows for computations on fully two-dimensional systems with \( W = L \). This enables a more accurate characterization of the scaling behavior and fluctuations across the entire range of \( p \), including the critical and crossover regimes.

It is useful to define two types of averages:  
\begin{enumerate}
\item the average over the distribution of uncut radii, denoted by $\langle \cdot \rangle$. The geometry of the percolating cluster is fixed. Note that in our model, the distribution of uncut radii depends on $p$, since the largest radii are systematically removed.
\item the total average, denoted by $\mathbb{E}(\cdot)$, which includes both the average over the distribution of radii and the average over the possible percolating clusters when $p \ge p_c$. 
\end{enumerate}

The optimization algorithms privilege paths that balance two competing criteria: selecting throats with large radii (low thresholds) and minimizing the total path length. This competition gives rise to different scaling regimes for the path length $\ell$ and the pressure threshold $\Delta P_0$.

In Fig.~\ref{fig:Energy_Q0}, we present numerical results for the mean and standard deviation of \( \Delta P_0 \) as a function of \( p \). Their scaling behaviors in the large system limit can be summarized by the following expressions:
\begin{align}
\Delta P_0(p) &= c_p L + \chi_p L^{1/3} && \text{for } p > p_c \tag{12a} \label{eq:P0_pneqpc} \\
\Delta P_0(p_c) &= \chi_{p_c} L^{1.13} && \text{for } p = p_c \tag{12b} \label{eq:P0_pc}
\addtocounter{equation}{1}
\end{align}
For \( p > p_c \), the scaling of \( \Delta P_0 \) follows that of the directed polymer Eq.~\eqref{eq:KPZ}: it is a self-averaging quantity with an extensive deterministic contribution and a sub-extensive stochastic term.
The coefficient \( c_p \) represents the deterministic pressure gradient threshold. 

At the percolation threshold \( p = p_c \), \( \Delta P_0(p_c) \) becomes super-extensive and fully stochastic. It is proportional to the chemical length of the percolation cluster, which scales as \( \ell_{\mathrm{chem}} \propto  L^{D_\mathrm{min}} \) with fractal dimension \( D_\mathrm{min} \simeq 1.13 \). Notably, both the chemical length and \( \Delta P_0 \) at criticality are not self-averaging: in the large-\( L \) limit, the fluctuations scale in the same way as the mean.  

Near the percolation threshold \( p_c \), the deterministic part of \( \Delta P_0 \) diverges as:
\begin{equation}
c_p \sim |p-p_c|^{-\theta}.
\end{equation}
The exponent $\theta$ obeys to a scaling relation ~\cite{barma85} noting that $\Delta P_0$ exhibits a size-dependent crossover. For system sizes smaller than the percolation correlation length \( \xi \propto (p - p_c)^{-\nu} \), with \( \nu = 4/3 \), \( \Delta P_0 \) displays the scaling behavior of Eq.~\eqref{eq:P0_pc} characteristic of the percolation point, namely \( \Delta P_0 \sim L^{D_\mathrm{min}} \) with \( D_\mathrm{min} \simeq 1.13 \). For larger systems \( L \gg \xi \), the effect of criticality is lost and \( \Delta P_0 \) crosses over to the directed polymer regime described by Eq.~\eqref{eq:P0_pneqpc}, with an extensive mean and \( L^{1/3} \) fluctuations.

\begin{figure*}[ht!]
    \centering
    \includegraphics[width=0.4\hsize]{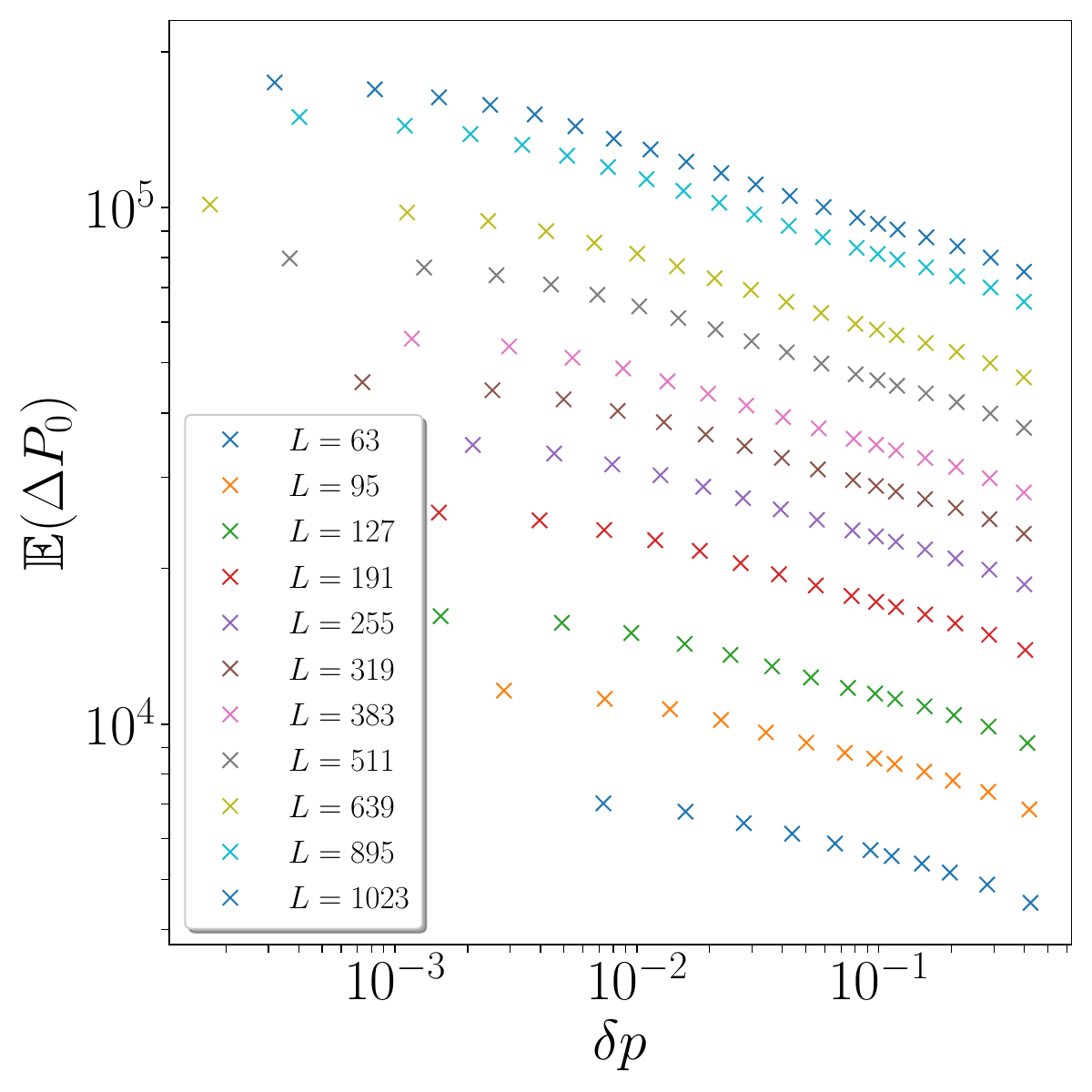}
    \includegraphics[width=0.4\hsize]{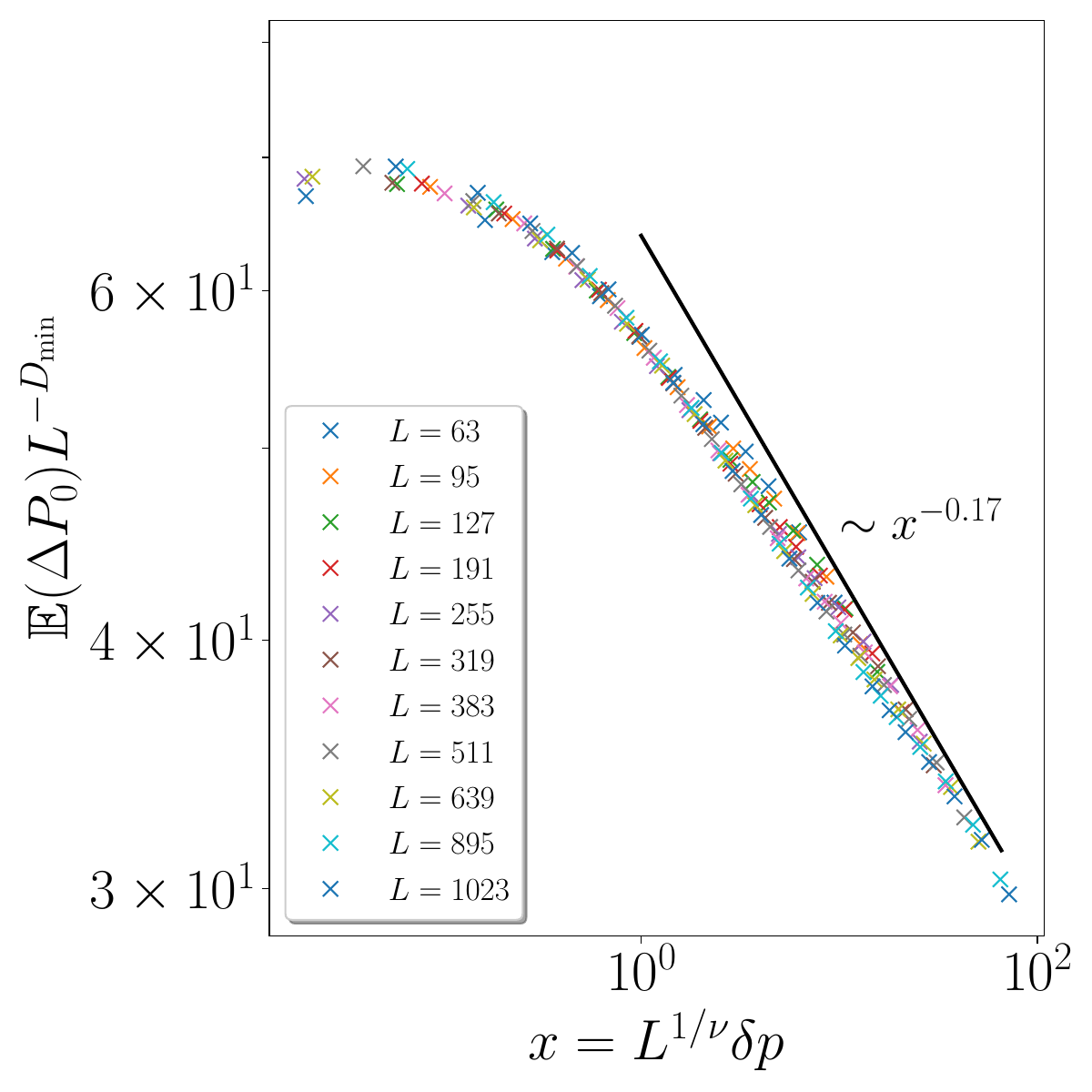}
    \caption{\textbf{Left:} Critical pressure threshold $\Delta P_0$ as a function of $\delta p = p- p_c$ for multiple system sizes $L$. Averages were performed for each value of $\delta p$ and $L$ over $300$ realizations of the disorder.
        \textbf{Right:} 
        Collapse using Eq. \eqref{eq:DeltaP0_scaling}. The power-law decay $x^{-\theta}$, with $\theta = 0.17$ is indicated by a solid line.
    }
    \label{fig:DP0_vicsek}
\end{figure*}

This behavior suggests the following scaling form:
\begin{equation}
    \Delta P_0 = L^{D_\mathrm{min}} f\!\left[ (p - p_c) L^{1/\nu} \right],
    \label{eq:DeltaP0_scaling}
\end{equation}
which we test through the data collapse shown in Fig.~\ref{fig:DP0_vicsek}. 
The argument of the scaling function $f$ is dimensionless. 
Exactly at the critical point $p = p_c$, we have $f(0) = \mathrm{const}$. 
Slightly above the critical point, for large $L$, the scaling variable $x = (p - p_c) L^{1/\nu}$ becomes large, and Eq.~\eqref{eq:DeltaP0_scaling} implies that $f(x) \propto x^{-\theta}$. 
On the other hand, since $\Delta P_0 \propto L$ must be recovered in this limit, the value of $\theta$ is then fixed as~\cite{barma85}
\begin{equation}
    \theta = \nu (D_\mathrm{min} - 1).
\end{equation}

\begin{figure*}[ht]
    \centering
    \includegraphics[width=0.32\hsize]{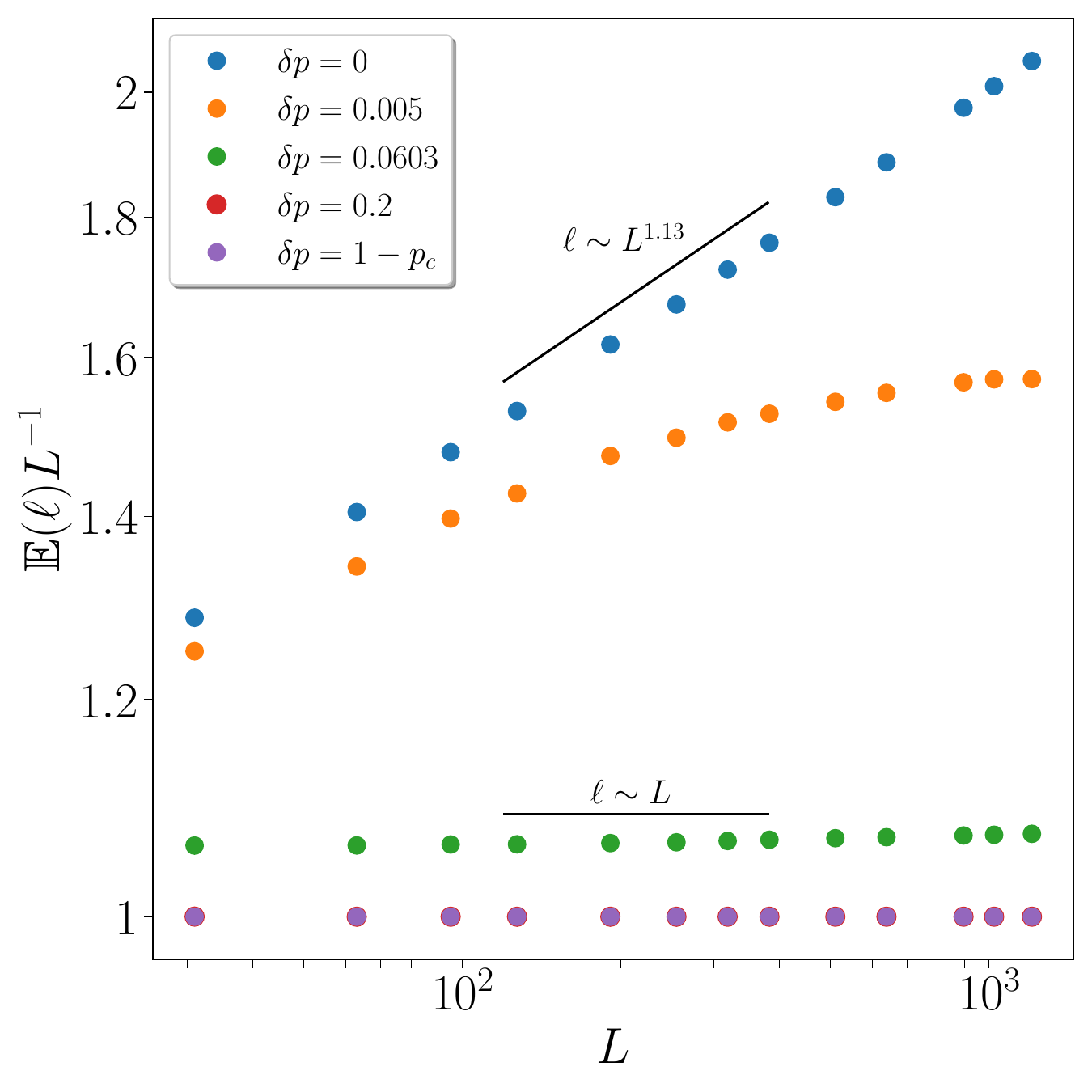}
    \includegraphics[width=0.32\hsize]{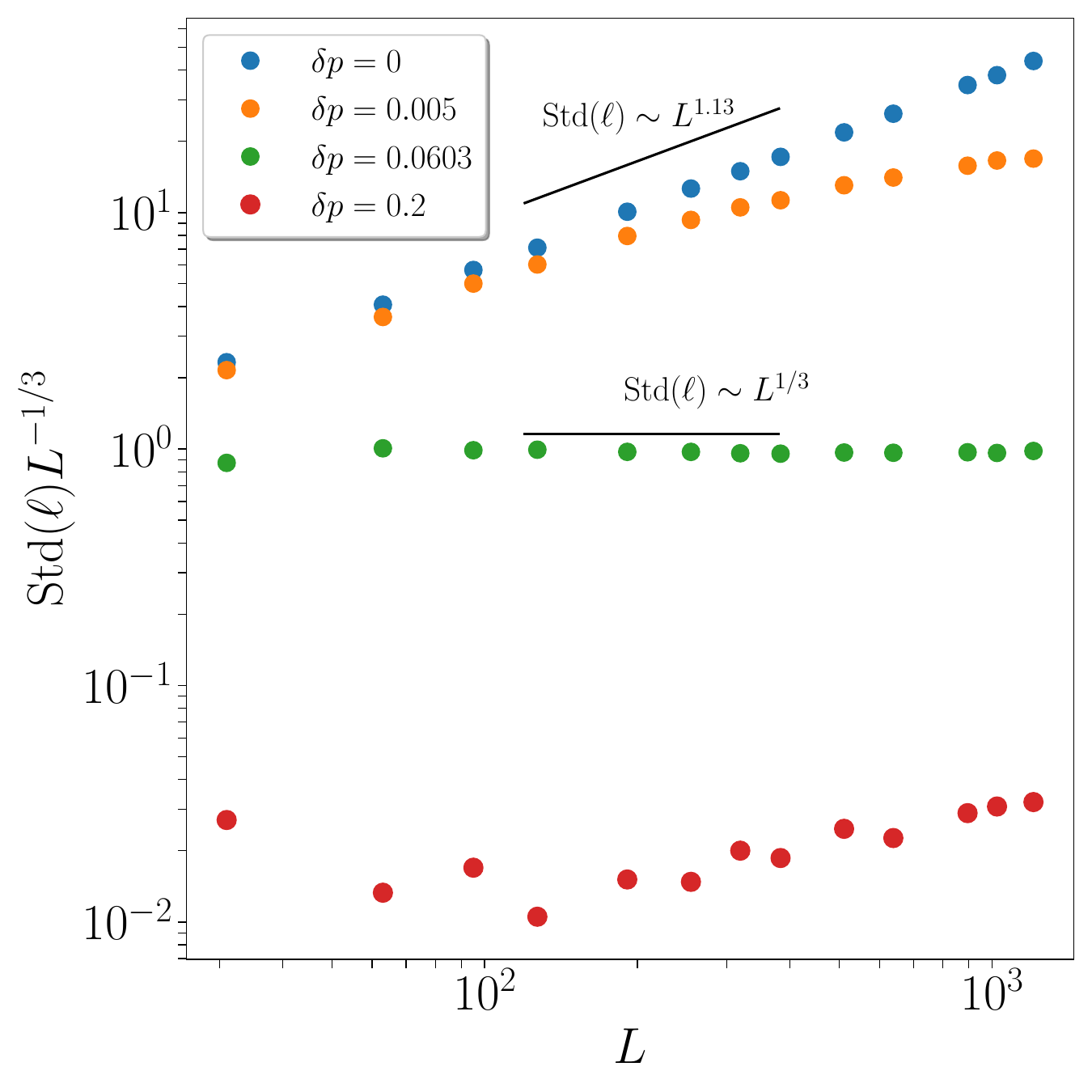}
    \includegraphics[width=0.32\hsize]{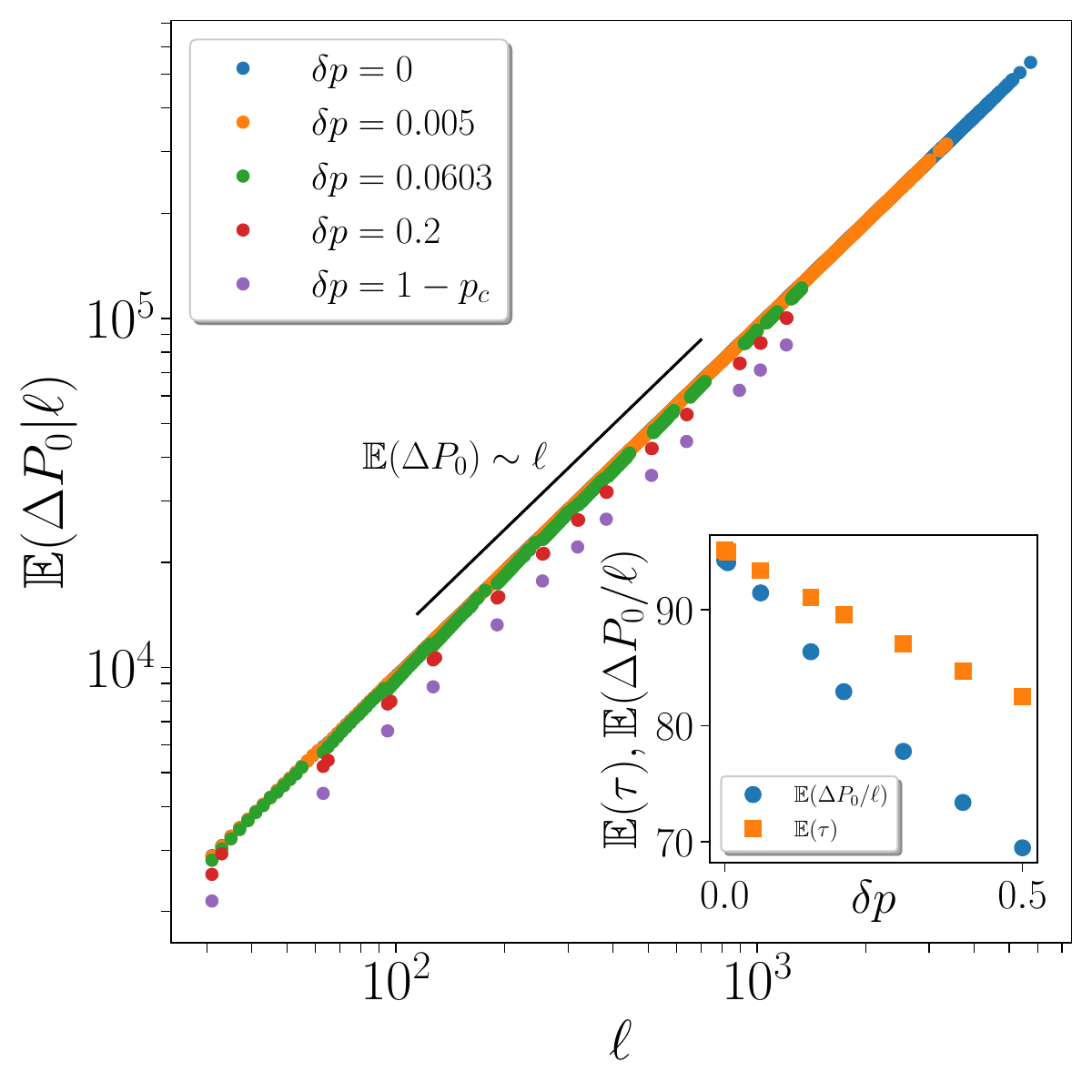}
    \caption{
        \textbf{Left:} Average length \( \ell \) of the first open path as a function of the system size \( L \) on a log-log scale.
        \textbf{Middle:} Standard deviation of \( \ell \) versus \( L \).
        \textbf{Right:}   $\Delta P_0$ as a function of the path length $\ell$. To reduce fluctuations, we average $\Delta P_0$ over different realizations with the same path length. The averages are computed from 10\,000 samples.
        Inset: Average local threshold along the selected path (blue) and average local threshold over the system (yellow). Owing to optimization, the former is smaller than the latter. Only at $\delta p = 0$ do the two become comparable, as the optimization is less effective.
    }
    \label{fig:ell_Q0}
\end{figure*}

\subsection{Non-directed Polymers in Random Media: Length of the Channel and Permeability}

The scaling of \( \Delta P_0 \) and its fluctuations with the system size \( L \) is closely related to the length \( \ell \) of the non-directed paths, as shown in Fig.~\ref{fig:ell_Q0}. Two distinct regimes can be identified: \( p > p_c \) and the critical point \( p = p_c \).

\begin{itemize}
    \item \textbf{\( p > p_c \):} In this regime, both \( \Delta P_0 \) and \( \ell \) are self-averaging and extensive. They are characterized by a deterministic part proportional to the system size \( L \) and subextensive fluctuations. The path length \( \ell \) is larger than \( L \) at all values of $p$ (Fig.~\ref{fig:ell_Q0}, Left). The directed percolation threshold \( p_d \) has no observable effect on the scaling of either \( \Delta P_0 \) or \( \ell \). This behavior arises from the competition between two equally significant contributions: selecting bonds with low local thresholds (large radii) and minimizing the total path length. Both effects carry comparable weight, leading to paths that are slightly elongated but composed of locally favorable links. As a result, the fluctuations of both \( \ell \) and \( \Delta P_0 \) scale as \( L^{1/3} \), consistent with the KPZ universality class (Fig.~\ref{fig:ell_Q0}, Middle).
   
    \item \textbf{\( p = p_c \):} At the percolation threshold, the number of available paths is drastically reduced and the minimal-length path, i.e. the chemical path, is effectively selected. In this case, the radii along the path are no longer large but rather typical, reflecting the fact that at \( p_c \) the only optimization criterion is minimizing the total length. The statistical properties of both \( \ell \) and \( \Delta P_0 \) are governed by the sample-to-sample fluctuations of the chemical length, which scales as \( \ell \sim L^{D_\mathrm{min}} \) with fractal dimension \( D_\mathrm{min} \simeq 1.13 \). Accordingly, \( \Delta P_0 \) follows the same anomalous scaling (Fig.~\ref{fig:ell_Q0}, Left and Middle).
\end{itemize}

As shown in Fig.~\ref{fig:ell_Q0} (Right), \( \Delta P_0 \) remains proportional to the total path length \( \ell \) across all values of \( p \), following a robust linear relationship. In the inset of Fig.~\ref{fig:ell_Q0} (Right), we compute, for each realization, the ratio $\Delta P_0/\ell$, which represents the average local threshold along the selected path. This ratio increases as $p$ approaches $p_c$, reflecting the reduced efficiency of the optimization mechanism. At $p_c$, it converges toward $\langle \tau \rangle$, which itself depends on $\delta p = p - p_c$, since removing the largest radii corresponds to eliminating the smallest values of $\tau$.  

From Eq.~\eqref{eq_permeability}, the low-flow-rate permeability is the inverse of the sum of the $\ell$ hydraulic conductances $\sigma_{ij}$ selected by the first open channel. Therefore, we expect
\begin{equation}
\kappa_0 = \kappa_{\mathrm{eff}}(\Delta P_0) \sim 1/\ell \sim 1/\Delta P_0 \, .
\end{equation}

Once again, two scaling regimes with system size $L$ can be anticipated:
\begin{itemize}
    \item \textbf{\(p > p_c\) :} In this regime, $\Delta P_0$ obeys Eq.~\eqref{eq:P0_pneqpc}. As a result, the deterministic part of the permeability scales as $L^{-1}$, while the standard deviation scales as $L^{-5/3}$.
    \item \textbf{\(p = p_c\) :} In this regime, $\Delta P_0$ obeys Eq.~\eqref{eq:P0_pc}. As a consequence, the permeability is non-self-averaging and both the mean and the fluctuations scale as $L^{-D_\mathrm{min}}$.
\end{itemize}

\begin{figure*}[ht]
    \centering
    \includegraphics[width=0.4\hsize]{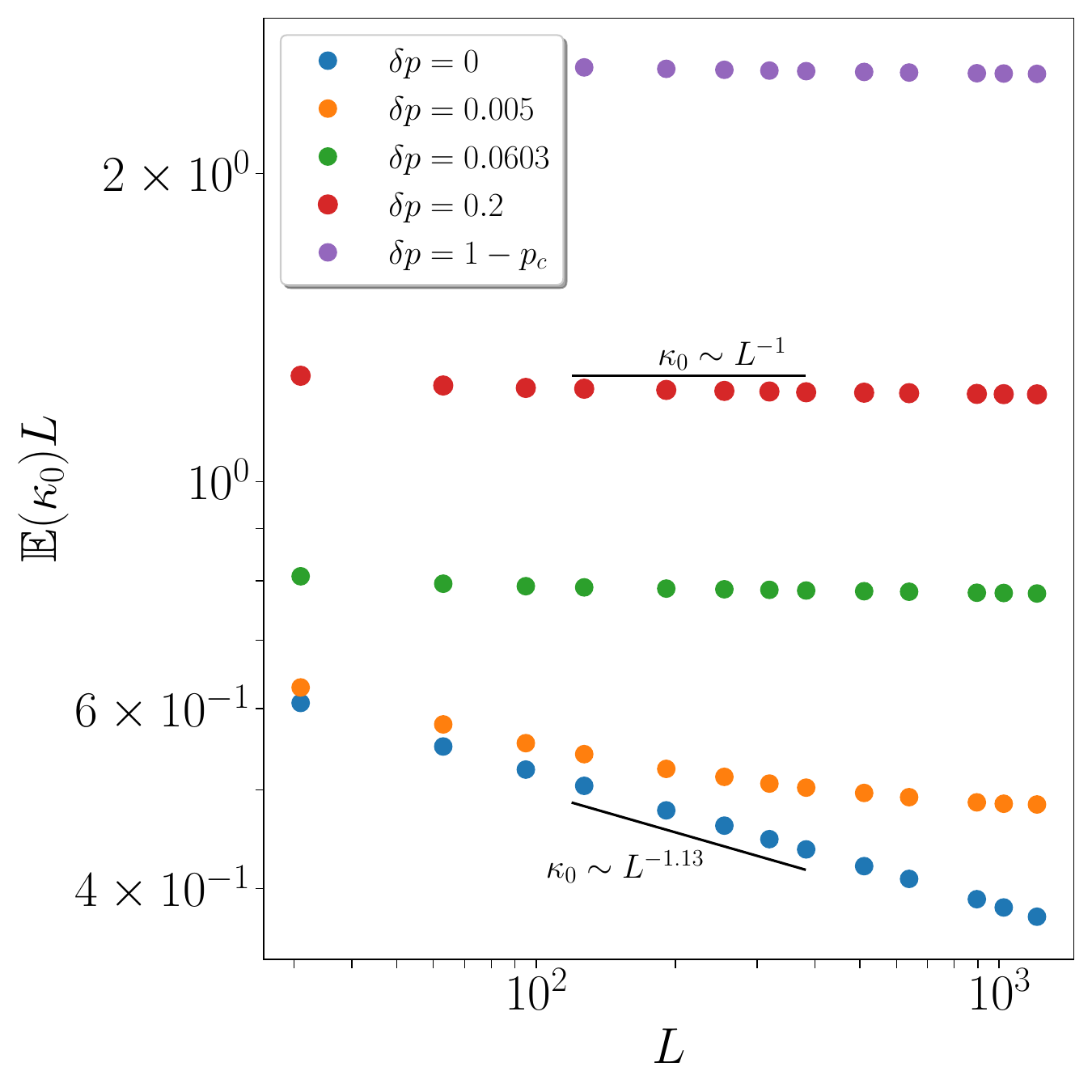}
    \includegraphics[width=0.4\hsize]{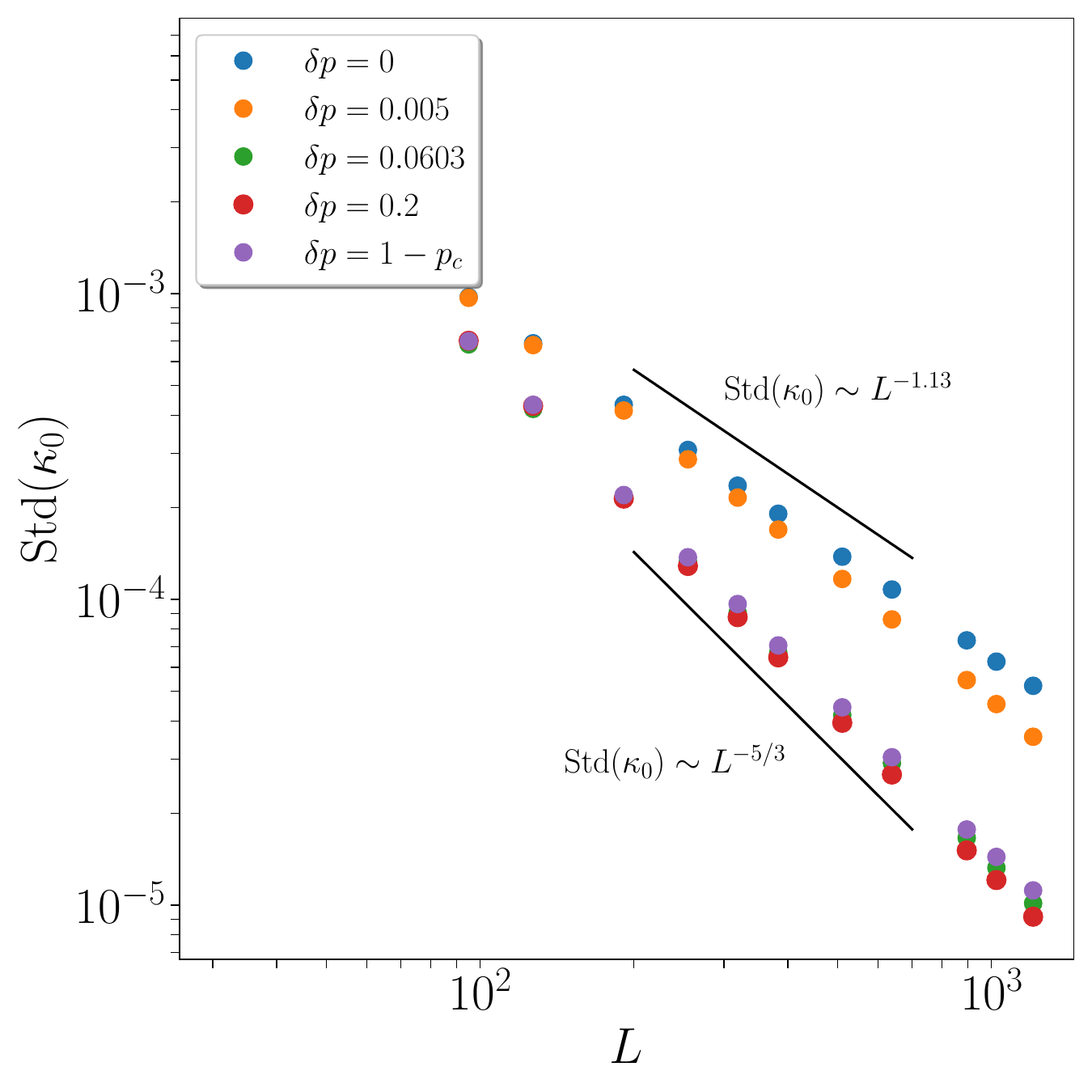}
    \caption{
        \textbf{Left:} Average effective permeability \( \kappa_0 \) as a function of the system size \( L \), at fixed distance from the percolation threshold \( \delta p = p - p_c \).
        \textbf{Right:} Standard deviation of \( \kappa_0 \) versus \( L \). Averages are performed over 10,000 disorder realizations.
    }
    \label{fig:kappa_0}
\end{figure*}

This behavior is clearly confirmed numerically; see Fig.~\ref{fig:kappa_0}.  

\section{\label{sec:high_Q} Large Flow Rate Limit $Q \to \infty$}

As in the low--flow--rate regime, the most interesting behavior in the high--flow--rate limit also occurs at the percolation threshold. Above this threshold, a finite correlation length $\xi$ emerges. For systems of size smaller than $\xi$, the behavior resembles that of the critical percolation cluster. For larger systems, the behavior crosses over to that already studied in the fully connected case $p=1$: both the effective permeability and the apparent critical pressure become self-averaging. The former becomes independent of the system size, while the latter scales linearly with $L$. In the following, we focus on the case $p = p_c$ and on the percolating cluster generated using Newman and Ziff’s algorithm \cite{newman01}.

\begin{figure*}[ht]
    \centering
    \includegraphics[width=0.4\hsize]{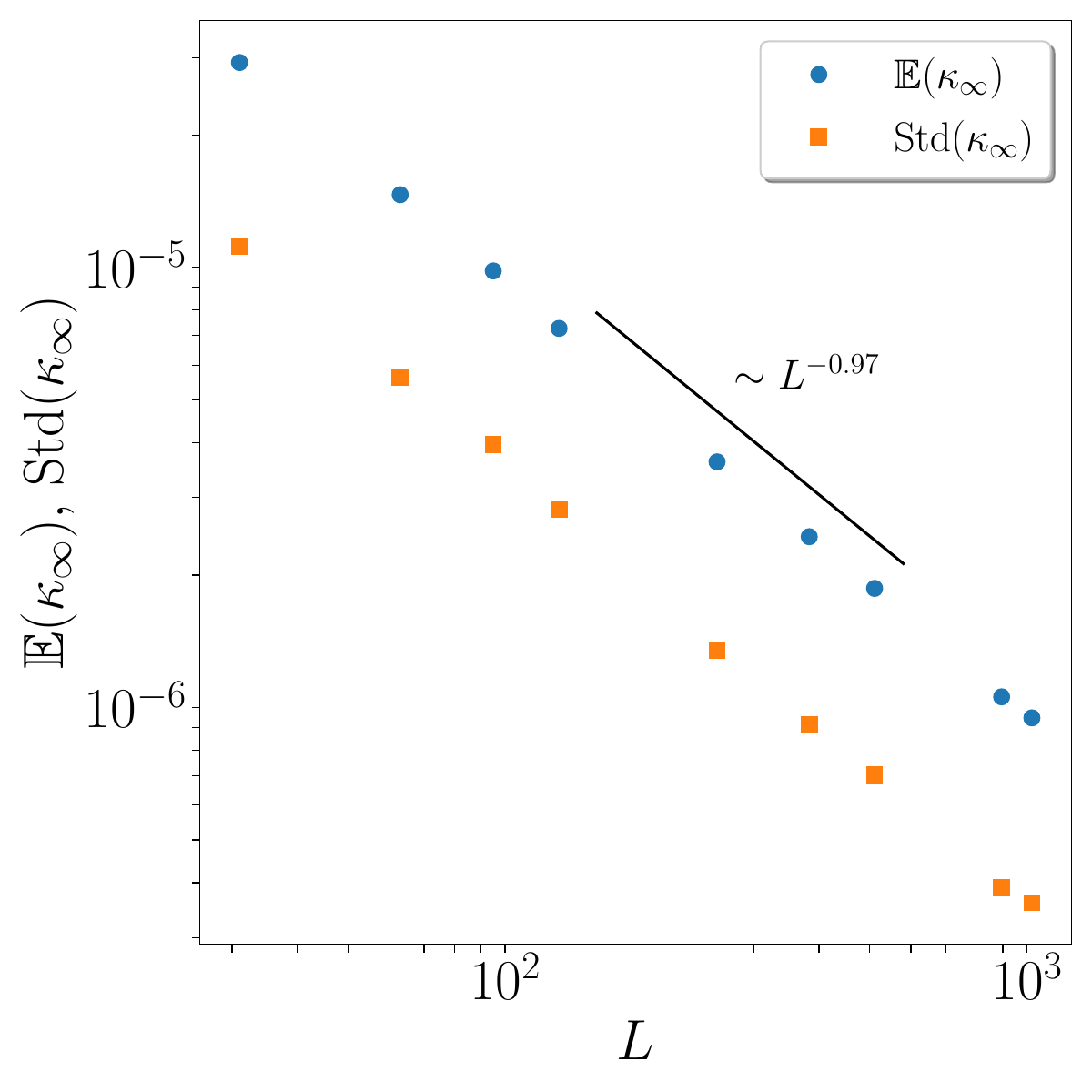}
    \includegraphics[width=0.4\hsize]{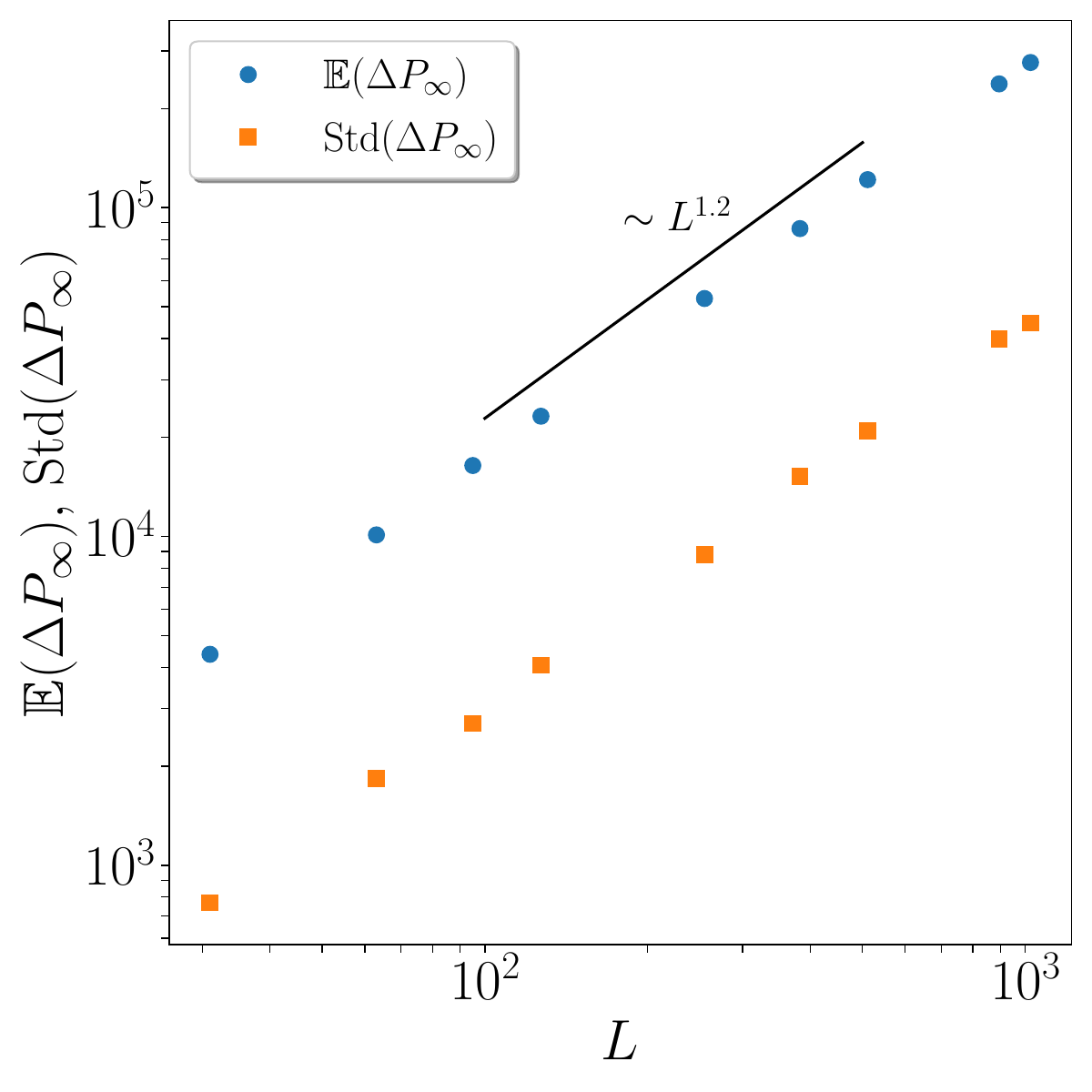}
\caption{\label{fig:kappa_infty}
Results for large flow rate at $p = p_c$, obtained by averaging over $N = 1000$ samples.  
\textbf{Left:} Mean permeability and its fluctuations. The permeability decays to zero, exactly as in the Newtonian case, with $L^{-t/\nu} \sim L^{-0.97}$. Note that self-averaging is lost: mean and fluctuations scale in the same way.  
\textbf{Right:} Mean offset and its fluctuations. The offset is proportional to the mean length of a streamline : $\Delta P_\infty \propto L^{D_\tau}$ with $D_\tau \approx 1.2$. The self-averaging is lost as the fluctuations scale as the mean. 
}
    \label{fig:k_infty}
\end{figure*}

In Fig.~\ref{fig:k_infty} we present the scaling of the permeability $\kappa_\infty$ and of the pressure offset $\Delta P_\infty$. At the percolation threshold $p=p_c$, sample-to-sample fluctuations associated with the structure of the backbone dominate once again. In particular, both the average and standard deviation of the permeability decay to zero in the same way as for Newtonian flow at the percolation point:
\begin{equation}
    \mathbb{E}(\kappa_\infty(p=p_c)) \sim L^{-t/\nu},
    \quad \mathrm{Std}(\kappa_\infty(p=p_c)) \sim L^{-t/\nu},
\end{equation}
where $\nu$ is the correlation-length exponent introduced above ($\nu=4/3$ in 2D), and $t$ is the conductivity exponent. Its value is not known analytically but has been determined numerically with good accuracy ($t \simeq 1.31$ in 2D)~\cite{Stauffer,kirkpatrick73,Stinchcombe76}. Hence, as for other quantities at $p_c$, the permeability is not self-averaging.
The same holds for $\Delta P_\infty$ (Figure \ref{fig:k_infty}, Right).

In the following, we give a physical interpretation of the behavior of the gradient 
\( \Delta P_\infty / L \simeq L^{0.2} \) 
and show that it scales in the same way as the tortuosity in the Newtonian limit of the critical percolation backbone \cite{lee99, berg24}, 
\( \mathcal{T} \sim L^{D_{\tau}-1} \), 
with \( D_\tau \simeq 1.2 \). 
The tortuosity is defined as the ratio between the average length of the streamlines within the backbone and the system size \( L \).

\begin{figure}[ht!]
    \centering
    \includegraphics[width=0.7\hsize]{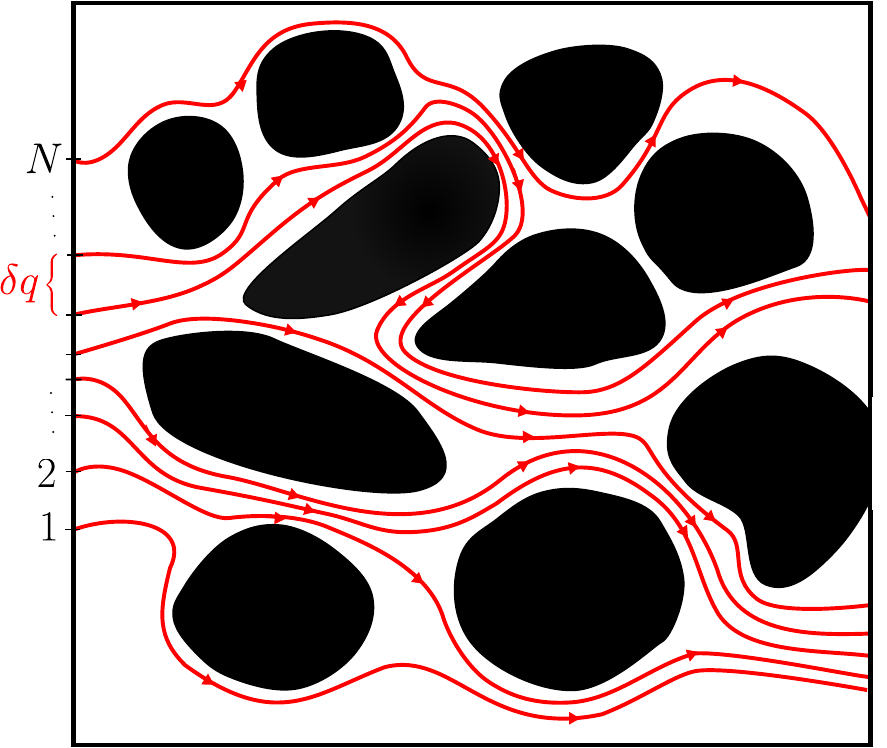}
    \caption{Illustration of the streamlines at the grain level.}
    \label{fig:streamline}
\end{figure}

To define the streamlines, it is useful to return to the microscopic description of the porous medium, 
composed of solid grains between which the fluid flows with a total flow rate \( Q \) (see Fig.~\ref{fig:streamline}). 
Streamlines are defined as iso-contours of the stream function. 
We select a set of \( N \) streamlines such that the volumetric flow rate between two consecutive streamlines is a constant increment \( \delta q \). 
By construction, the total flow rate satisfies \( Q = N \, \delta q \). 
The tortuosity can be written as
\begin{equation}
    \mathcal{T} = \frac{1}{N} \sum_{\mathcal{S}=1}^{N} \frac{l_{\mathcal{S}}}{L},
    \label{eq:tortuosity}
\end{equation}
where the sum runs over all streamlines and \( l_{\mathcal{S}} \) is the length of streamline \( \mathcal{S} \). 
In the pore network model, this length can be approximated by summing the lengths of the throats traversed by the streamline. 

The sum in Eq.~\eqref{eq:tortuosity} can thus be recast as a sum over all throats of the pore network, 
weighted by the number of streamlines passing through each throat:
\begin{equation}
    \mathcal{T} = \frac{\delta q}{Q L} \sum_{\langle ij \rangle} l_{ij} n_{ij},
\end{equation}
where \( n_{ij} \) denotes the number of streamlines passing through throat \( \langle ij \rangle \). 
Since the throat length is \( l_{ij} = 1 \) and the local flow rate satisfies \( |q_{ij}| = n_{ij} \, \delta q \), we obtain:
\begin{equation}
    \mathcal{T} = \frac{1}{Q L} \sum_{\langle ij \rangle} |q_{ij}| 
    \simeq \frac{1}{\mathbb{E}(\tau)} \, \frac{\Delta P_{\infty}}{L},
\end{equation}
where, in the last equality, we use the definition given in Eq.~\eqref{eq:Pstar_def}. 
In the limit of high flow rate, the flow field is identical to that of a Newtonian fluid, 
and the tortuosity therefore corresponds to the Newtonian one.

\section{\label{sec:hom} Discussion and Effective Medium Picture}

The results obtained at the percolation threshold clearly show that the behavior in both limits $Q \to 0$ and $Q \to \infty$ is governed by the geometry of the given percolating cluster and is independent of the particular realizations of the uncut radii. It is therefore natural to conjecture that not only these limiting cases, but also the entire sequence of channels that successively open as the pressure gradient increases, are determined only by the geometric properties of the percolating cluster.

\begin{figure*}[ht!]
    \centering
    \includegraphics[width=0.85\hsize]{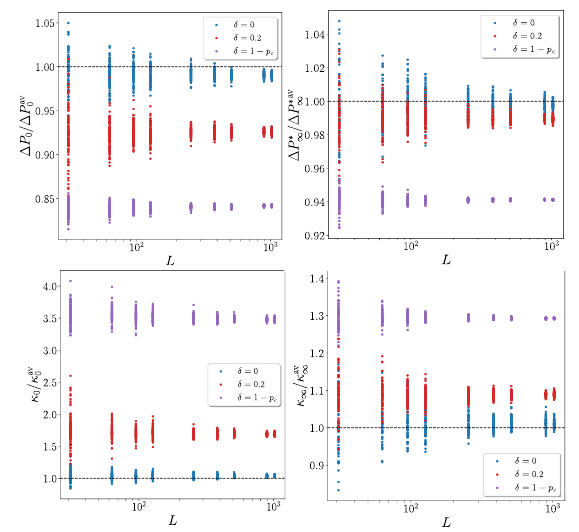}
    \caption{
            Test of the homogeneous effective-medium approximation for the flow-curve parameters $\Delta P_0$ (top left), $\kappa_0$ (bottom left), $\Delta P_\infty$ (top right), and $\kappa_\infty$ (bottom right). 
            Results are obtained from $100$ disorder realizations for various system sizes $L$ and values of $\delta p = p - p_c$.
    }
    \label{fig:hom_flow_parameters}
\end{figure*}

To test this hypothesis, we introduce an {\em effective homogeneous medium} in which the percolation lattice is fixed, but the uncut throats are no longer disordered. Instead, they are characterized by the average conductivity and the average threshold :
\begin{align}
    \sigma^{\text{av}} & = \langle \sigma \rangle =   \frac{\pi}{8\eta \langle r_{ij}^{-4} \rangle}
    \label{eq:sigma_av} \\
    \tau^{\text{av}}  & = \langle \tau \rangle = 2\Tau_c l  \langle  r_{ij}^{-1} \rangle
    \label{eq:tau_av}
\end{align}
Within this approximation, the geometry of the percolating cluster remains random, while the conductivity and the threshold are taken to be deterministic and equal to their average values. For this reason, we denote the observables obtained within this approximation using the superscript ``$\mathrm{av}$''. To assess the validity of this approximation, we first focus on the four parameters $\Delta P_0$, $\kappa_0$, $\kappa_\infty$, and $\Delta P_\infty$. In Fig.~\ref{fig:hom_flow_parameters}, we report, for each observable, the ratio between its value computed for a random realization of the uncut radii and the corresponding value obtained within the effective homogeneous medium approximation, evaluated on the same percolating cluster. At the percolation point $\delta p = 0$, the ratio approaches one as the system size grows while above the percolation point, the ratio is very different from one. As a result, the effective homogeneous medium approximation holds only at $p = p_c$ for large systems.

\begin{figure*}[ht!]
    \centering
    \includegraphics[width=0.4\hsize]{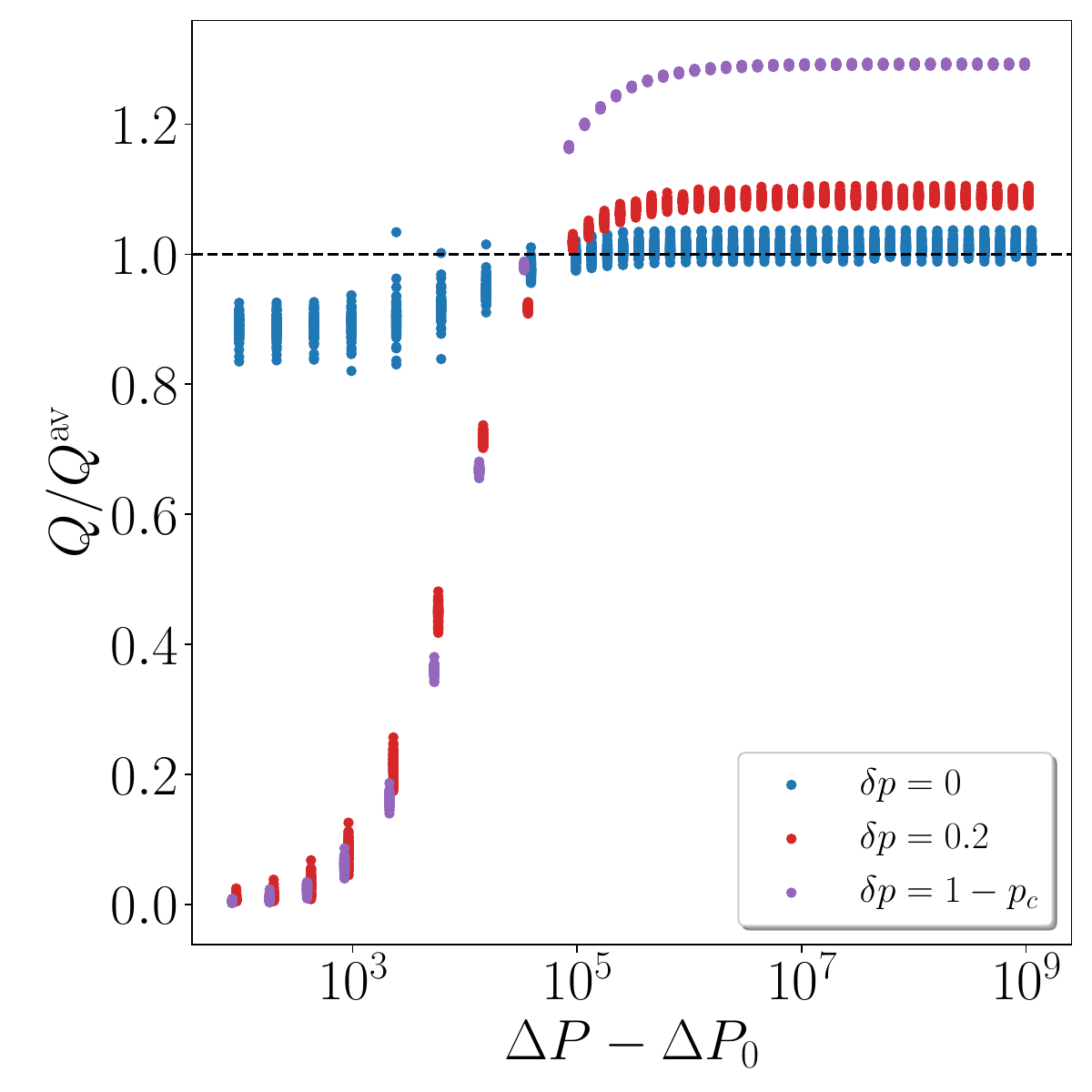}
    \includegraphics[width=0.4\hsize]{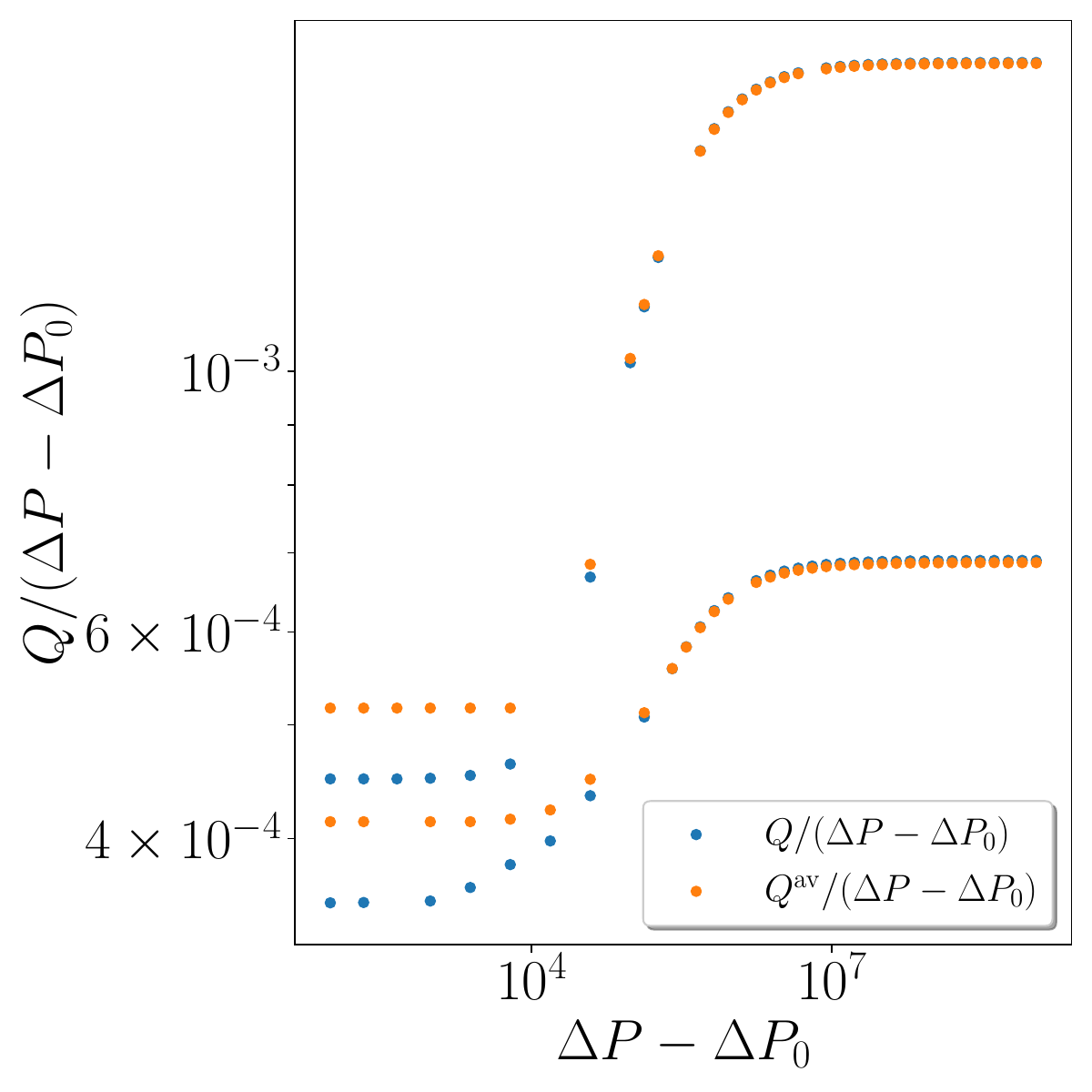}
    \caption{
        \textbf{Left:}
        Ratio of the flow rates $Q$ and $Q_\mathrm{av}$ for a system of size $L = 1023$, as a function of $\Delta P - \Delta P_0$, and for multiple values of $\delta p = p - p_c$. 
        Here, $Q$ denotes the flow rate obtained for a given realization of the disorder, while $Q^\mathrm{av}$ corresponds to the flow rate computed within the effective homogeneous medium approximation in which the  percolation lattice is fixed.
        \textbf{Right:} 
        Evolution of the flow rates with $\Delta P - \Delta P_0$ : two realizations of the disorder at the percolation threshold (orange) are compared with the prediction $Q^\mathrm{av}$ (blue) from the effective homogeneous-medium approximation. Only at the percolation point do the ratios approach one, indicating the validity of the effective homogeneous medium approximation.
    }
    \label{fig:Q_vs_hom}
\end{figure*}

In Fig.~\ref{fig:Q_vs_hom}, we assess the validity of the effective homogeneous medium approximation for the full flow curve. As before, above the percolation threshold ($\delta p > 0$), the approximation breaks down, and the ratio $Q / Q^{\mathrm{av}}$ deviates significantly from unity (Fig. \ref{fig:Q_vs_hom} Left). At the percolation point, however, the approximation performs well, particularly at large flow rates. At low flow rates, it slightly overestimates the flow (see Fig. \ref{fig:Q_vs_hom} Right). This behavior can be understood from the fact that, in this regime, the flow localizes along the shortest path, referred to as the ``chemical path''. In a finite system, the chemical path may be degenerate, and the disorder in the radii lifts this degeneracy. Within the effective homogeneous medium approximation instead, all chemical paths are treated as equivalent; therefore, at the threshold, many such paths become active simultaneously, increasing the flow rate. Another consequence of being at the percolation point is that the intermediate quadratic regime known when $p=1$ \cite{roux87,talon13b} vanishes (see Appendix \ref{app:regimes}).

\section*{Conclusion}

In this work, we have investigated the flow of a Bingham yield-stress fluid through an \emph{unsaturated} porous medium, modeled as a two-dimensional disordered pore network of length \(L\) and width \(W\). In such systems, the presence of a non-wetting phase trapped within the pore space obstructs a fraction of the throats. This obstruction is modeled by removing links whose diameter exceeds the critical capillary threshold, so that the non-wetting fluid saturation corresponds to a fraction \(1-p\) of blocked links. This setting naturally creates a percolation problem: the yield-stress fluid must flow through a disordered medium in which some flow paths are inaccessible. The combined effect of Bingham rheology and structural disorder qualitatively alters Darcy’s law: the flow rate \(Q\) becomes nonlinear in the applied pressure gradient and vanishes below a critical value \(\Delta P_0/L\). Above this threshold, the system exhibits an effective permeability \(\kappa_{\mathrm{eff}}\) and an offset \(\Delta P^*\), both of which increase with the applied pressure.

Unlike previous approaches that restricted the flow to directed paths, here we allow for all possible paths—directed and non-directed—and analyze how the flow rate decreases when a fraction \(p\) of the throats is blocked. Two distinct regimes emerge, separated by the percolation threshold \(p_c\).

\paragraph{The regime \(p>p_c\).}
When the network remains globally connected, both the critical pressure gradient \(\Delta P_0/L\) and the large-flow offset \(\Delta P_\infty/L\) take finite values in the thermodynamic limit.
The low flow rate permeability vanishes as
\begin{equation}
\kappa_0 \sim 1/L ,
\end{equation}
while at large flow rates the effective permeability \(\kappa_\infty\) approaches a finite constant, equal to the Newtonian permeability of the same network.
All quantities are self-averaging, and the finite-size fluctuations of the flow parameters  are subleading, with exponents consistent with the KPZ universality class at low flow rates.

\paragraph{The critical regime \(p=p_c\).}
At the percolation threshold, the flow is constrained to the percolating backbone.
In this regime, the average values and fluctuations of the flow parameters scale identically with system size, so that the system is not self-averaging.
At low flow rates, the geometry of the flow is controlled by the chemical length $\ell_\mathrm{chem} \sim$ \(L^{D_\mathrm{min}}\) of the backbone, leading to the scaling relations
\begin{equation}
\Delta P_0/L \sim L^{D_\mathrm{min}-1}, \qquad
\kappa_0 \sim 1/L^{D_\mathrm{min}}.
\end{equation}
At large flow rates, the relevant length scales are controlled by the conductivity exponent $t$ and the tortuosity exponent $\mathcal{T} \simeq L^{D_\tau-1}$ , yielding
\begin{equation}
\kappa_\infty \sim 1/L^{t/\nu}, \qquad 
\Delta P_\infty/L \sim L^{D_\tau-1}.
\end{equation}
Here \(D_\mathrm{min} \simeq 1.13\) is the fractal dimension of the chemical length, \(\nu = 4/3\) is the correlation-length exponent, \(D_\tau \simeq 1.2\) characterizes the average streamline length, and \(t =1.31 \) is the conductivity exponent. For a given realization of the disorder, the system is well approximated by a homogeneous effective medium and the flow is effectively controlled by the geometry of the available flow paths.

\paragraph{Crossover for \(p \gtrsim p_c\).}
The transition between the two regimes can be described by the scaling forms
\begin{align}
    \frac{\Delta P_0}{L} & \sim (p-p_c)^{-\nu (D_\mathrm{min}-1)}, \nonumber \\ 
    \frac{\Delta P_\infty}{L} & \sim (p-p_c)^{-\nu (D_\tau-1)}, \nonumber \\
    \kappa_\infty & \sim (p-p_c)^{t}. \nonumber
\end{align}
These relations quantitatively capture the evolution from the critical, non-self-averaging regime at \(p=p_c\) to the homogeneous, self-averaging behavior at \(p>p_c\).

In summary, this work develops a framework for describing yield-stress flow in disordered porous media, showing how the interplay between geometric disorder and local plastic thresholds govern both the onset of flow and its scaling properties. The resulting picture links rheological nonlinearities with percolation geometry and with scaling features characteristic of depinning and KPZ universality.

This establishes a quantitative link between yield-stress rheology and critical transport on percolation networks, and suggests that extensions to three-dimensional systems, correlated disorder, or time-dependent driving represent promising directions for future work.

\section{Acknowledgements} 

This work was partly supported by the Research Council of Norway through the INTPART program (project number 309139) and its Centers of Excellence funding scheme (project number 262644). AH also acknowledges funding from the European Research Council (Grant Agreement 101141323 AGIPORE).

\appendix

\section{Derivation of Eq.~\eqref{eq:nonlinear_darcy}}
\label{app:darcy}
To obtain Eq. \eqref{eq:nonlinear_darcy}, we need to write the equation for the total flow rate $Q$ in a convenient form.
To do that, we first  derive from Eq. \eqref{bingham_link_flow} the pressure difference of an open link:
\begin{equation}
    \delta P_{ij} = \frac{1}{\sigma_{ij}} q_{ij} + \tau_{ij} \frac{q_{ij}}{|q_{ij}|} \;\;\; \text{with} \;\;\; q_{ij}\neq 0.
\end{equation}
Multiplying by $q_{ij}$ , summing over every link and using the conservation of mass in each node, it follows:
\begin{equation}
 Q \Delta P = \sum_{\langle ij \rangle} (\frac{1}{\sigma_{ij}} q^2_{ij} + \tau_{ij} |q_{ij}|).
 \label{eq:dissip_link}
\end{equation}
From equation Eq.~\eqref{eq:dissip_link} we can then write the non-linear Darcy's law equation.

\section{Influence of percolation on the flow rate curve}
\label{app:regimes}
\begin{figure}[ht!]
    \centering
    \includegraphics[width=0.85\hsize]{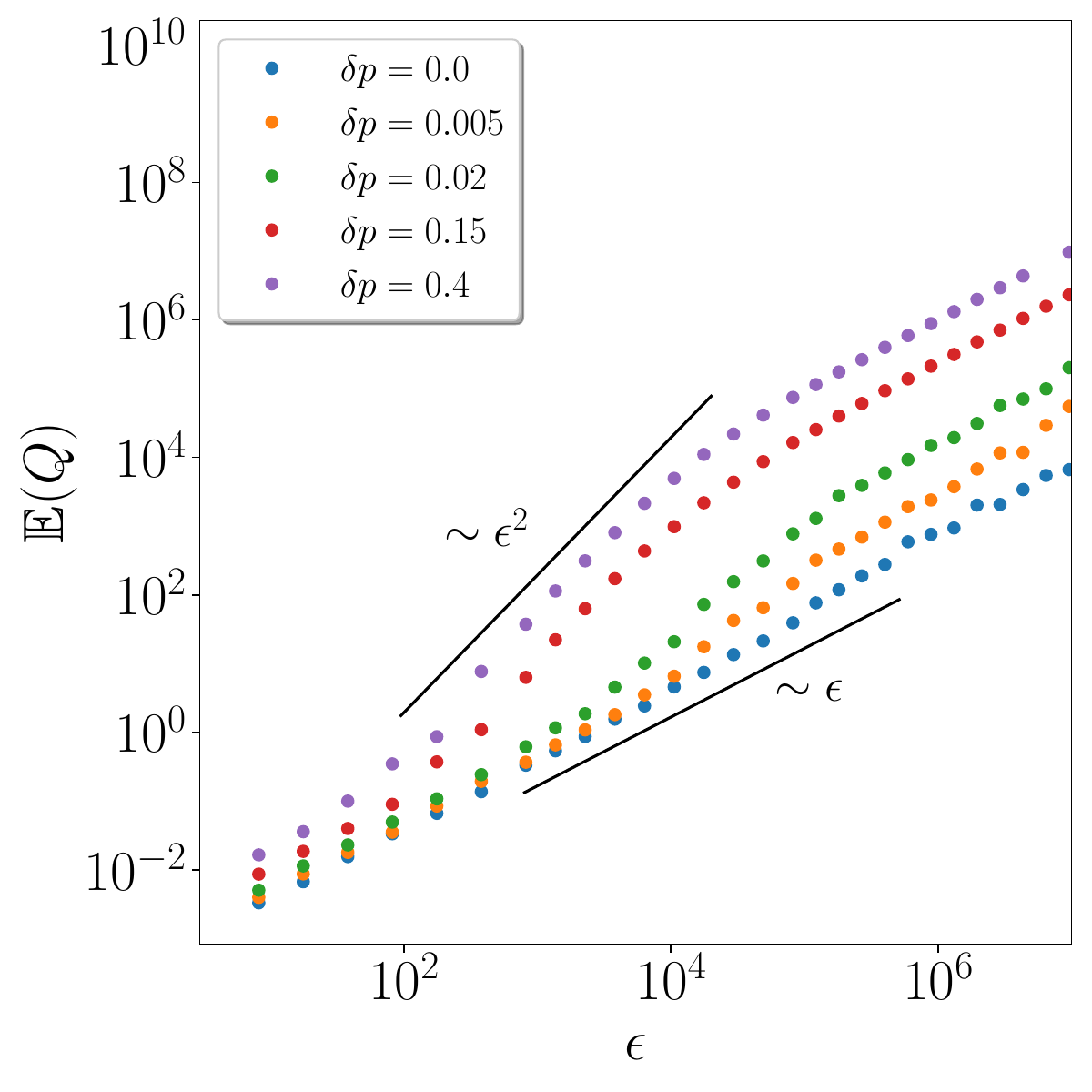}
    \caption{ Flow rate $Q$ as a function of the pressure drop difference $\epsilon = \Delta P - \Delta P_0$ for a system of size $L = 1023$ and for multiple values of $\delta p = p - p_c$. The flow rate was averaged over 5 realizations.
    }
    \label{fig:flow_rate_transition}
\end{figure}

In the main text, we concentrated on the two limiting regimes. In this appendix, we explore more closely how percolation modifies the flow rate $Q$ when plotted as a function of the pressure-drop excess $\Delta P - \Delta P_0$, with particular attention to the fate of this intermediate quadratic regime. When $p = 1$, i.e., when no links are removed from the network, three distinct flow regimes are known to occur \cite{roux87,talon13b}. The first arises when $\Delta P$ is only slightly above the critical pressure threshold $\Delta P_0$, so that the flow is confined to a single path. Since the local flow rate in Eq.~\eqref{bingham_link_flow} is linear in the pressure drop, the global flow rate $Q$ is also linear in $\Delta P$ in this regime. At the opposite limit, when $\Delta P \gg \Delta P_0$, all paths carry flow and the Newtonian regime is recovered, where $Q$ again varies linearly with $\Delta P$. In the intermediate regime between these two limits, however, the flow rate follows a quadratic law, $Q \sim (\Delta P - \Delta P_0)^2$. This behavior results from the incremental increase of the effective permeability as the pressure drop is increased and new paths start flowing. We present on in Fig.~\ref{fig:flow_rate_transition} the case $p < 1$. In the limit $p \rightarrow 1$, we recover the expected quadratic dependence of $Q$ on $\Delta P - \Delta P_0$. As the percolation point is approached, however, this quadratic regime vanishes: the increase in permeability becomes negligible, leaving only a linear regime over the entire flow curve. We attribute this behavior to the scarcity of accessible flow paths near the percolation point: the few paths that remain percolating are highly tortuous and therefore contribute only marginally to the global flow rate.

 \bibliography{biblio}

\end{document}